# Factors resisting protein adsorption on hydrophilic/hydrophobic self-assembled monolayers terminated with hydrophilic hydroxyl groups

Dangxin Mao, Yuan-Yan Wu[*], and Yusong Tu[*]

*College of Physics Science and Technology, Yangzhou University, Jiangsu 225009, China*

[*]Corresponding author: yywu@yzu.edu.cn; ystu@yzu.edu.cn

The hydroxyl-terminated self-assembled monolayer (OH-SAM), as a surface resistant to protein adsorption, exhibits substantial potential in applications such as ship navigation and medical implants, and the appropriate strategies for designing anti-fouling surfaces are crucial. Here, we employ molecular dynamics simulations and alchemical free energy calculations to systematically analyze the factors influencing resistance to protein adsorption on the SAMs terminated with single or double OH groups at three packing densities ($\varSigma$ = 2.0 nm$^{-2}$, 4.5 nm$^{-2}$, and 6.5 nm$^{-2}$), respectively. For the first time, we observe that the compactness and order of interfacial water enhance its physical barrier effect, subsequently enhancing the resistance of SAM to protein adsorption. Notably, the weak spatial hindrance effect of SAM leads to the embedding of protein into SAM, resulting in a lack of resistance of SAM towards protein. Furthermore, the number of hydroxyl groups per unit area of double OH-terminated SAM at $\varSigma$ = 6.5 nm$^{-2}$ is approximately 2 to 3 times that of single OH-terminated SAM at $\varSigma$ = 6.5 nm$^{-2}$ and 4.5 nm$^{-2}$, consequently yielding a weaker resistance of double OH-terminated SAM towards protein. Meanwhile, due to the structure of SAM itself, i.e., the formation of a nearly perfect ice-like hydrogen bond structure, the SAM exhibits the weakest resistance towards protein. This study will complement and improve the mechanism of OH-SAM resistance to protein adsorption, especially the traditional barrier effect of interfacial water.

## 1 Introduction

The adsorption of biological organic macromolecules, such as protein and glycoprotein, onto the underwater substrate is the primary step in the formation of marine biofouling.[1-3] Their adherence to the hull surface will increase surface roughness, consequent navigation resistance and fuel consumption, and more severe acceleration of corrosion,[4] leading to potential safety concerns and economic losses. Therefore, it is crucial for enhancing the anti-fouling properties of surfaces to reduce the probability of life-threatening incidents and operational costs. OH-SAMs are extensively utilized in anti-fouling research due to their excellent bioinertness and biocompatibility.

Experimental and theoretical researchers[5-13] revealed that OH-SAMs exhibited unique advantages in resistance to protein/peptide adsorption, which was ascribed to a physical barrier effect formed by a number of hydrogen bonds between interfacial water and OH-SAM with weak charge distribution in the headgroups. This physical barrier effect was traditional mechanistic explanation for OH-SAM resistance to protein adsorption. However, our previous study[14] found that SAM terminated with only hydrophilic OH groups exhibited different hydrophilic and hydrophobic characteristics at different packing densities. For such structures, is the effect of interfacial water on OH-SAM resistance to protein adsorption consistent? Does only interfacial water influence the resistance of OH-SAM to protein adsorption? These issues are necessary to explore for the rational design and precise customization of anti-fouling surfaces. Our recent study[15] has also proposed a new mechanism of resistance to protein adsorption, namely, the structure of SAM itself.

In this study, we utilize molecular dynamics (MD) simulations and alchemical free energy calculations to systematically investigate the factors affecting the resistance of single or double OH-terminated SAM to protein adsorption at three packing densities (2.0 nm$^{-2}$, 4.5 nm$^{-2}$, and 6.5 nm$^{-2}$). The aim is to enrich and improve the mechanism of OH-SAM resistance to protein adsorption, provide directional guidance for the rational design and precise customization of OH-terminated anti-fouling surfaces, and promote their applications in ship navigation and medical implants.

## 2 Models and methods

### 2.1 Models

The SAM consisted of five-carbon long alkyl chains, one end of which was terminated with single or double hydrophilic OH groups, and the model atoms at the other end were bound to the 111 surface (Figure 1). The chain numbers of SAMs were the same (440). The packing densities of SAMs were 2.0 nm$^{-2}$, 4.5 nm$^{-2}$, and 6.5 nm$^{-2}$, representing low, medium, and high densities, respectively. The water contact angles for SAMs terminated with two hydrophilic OH groups at the three packing densities were 35.6°, 82°, and 0°,[14] respectively, while those for SAMs terminated with the single OH group were 63.4°, 22°, and 0° (Figure S3 in the Supplementary Information), respectively. The calculation method for contact angles is consistent with our previous work.[14, 16, 17] The two OH-terminated SAM at $\varSigma$ = 4.5 nm$^{-2}$ with an 82° contact angle is defined as hydrophobic, and the other surfaces of the six systems were classified as hydrophilic.[14, 18, 19] The crystal structure of protein used in the simulations was taken from the Protein Data Bank (PDB ID: 5DUY).[20] As a mussel protein, it was widely used to study the SAM resistance to protein adsorption[21] due to strong adhesive property.[22, 23] The protein was initially placed 6 Å above the SAM. The protein-SAM complex was solvated with SPC/E[24] water molecules. Sodium and chloride ions were added to neutralize the system with a concentration of 0.15 M.

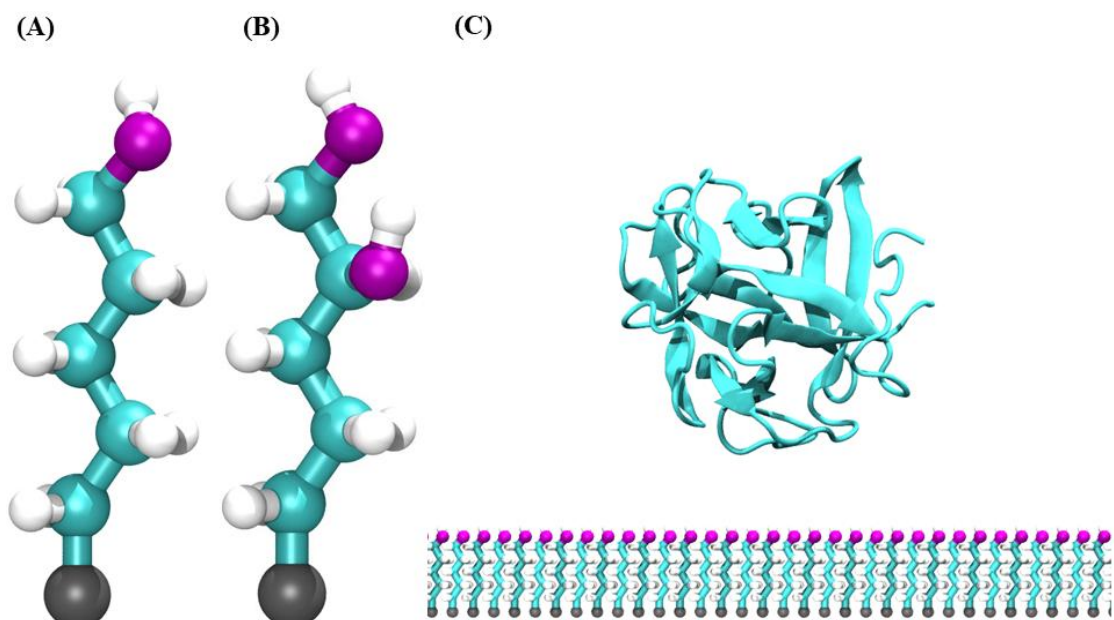

Figure 1. The initial structures of single alkyl chains terminated with an OH group (A) and two OH groups (B), respectively, and the initial structure of protein on the SAM (C). The protein is displayed using the NewCartoon mode, and the alkyl chains are represented in Licorice mode. The rest of the system is omitted for clarity. Color codes: hydroxyl groups, purple and white; main chain, cyan and white; and model atoms, gray.

**2.2 Equilibrium MD simulations**

MD simulations were performed using GROMACS 2019.6 package[25, 26] with an OPLS-AA force field.[27] The force field parameters for the alkyl chain were derived from our previous study.[14, 15] The particle-mesh Ewald (PME) algorithm[28, 29] was used to treat electrostatic interactions with a 1.2 nm cutoff radius. Likewise, a cutoff radius for the van der Waals interaction was 1.2 nm. Newton's equation of motion is integrated using the leap-frog algorithm with a 2 fs integration time step. Periodic boundary conditions were applied in all directions. A velocity rescaling thermostat was used to maintain the temperature at 300K. All bonds involving hydrogen atoms were constrained using the LINCS algorithm.[30] All systems were equilibrated in the canonical (*NVT*) ensemble for 300 ns, and the last 20 ns were used for data analysis. The coordinates were saved every 4 ps.

**2.3 Alchemical free energy calculations**

The last structures from MD simulations were used as the starting configurations for further alchemical simulations. The alchemical transformations were divided into two processes, one was the decoupling of protein from the water solution in the solvated protein system, and the other was the decoupling of protein from the solvated protein-SAM complex. The free energy differences (Δ*G*) were defined as the differences between the free energy changes of the decoupling of protein from the water solution in solvated protein system and those from the solvated complex,[31, 32] which was used to characterize the binding affinity of protein to the SAM. A coupling parameter, $\lambda$, ranging from 0 to 1 was introduced to alchemical transformations with an equally spaced value of 0.1 and 21 $\lambda$ windows. It was noticed that $\lambda = 0$ represented the initial state where the electrostatic and van der Waals interactions between protein and the surroundings were fully turned on, while $\lambda = 1$ represented the final state where there was no interaction between protein and the surroundings. The electrostatic interactions were first linearly scaled to 0, and then the van der Waals interactions were turned off using a soft-core potential[33] to avoid singularities. Langevin dynamics[34] were implemented for temperature regulation with a friction coefficient of 2 ps$^{-1}$ and a reference temperature of 300 K in the *NVT* ensemble. Each $\lambda$ window was equilibrated for 1 ns, and then was sampled for 1 ns. The free energy changes were estimated using the Multistate Bennett Acceptance Ratio (MBAR) method[35] implemented in the pymbar package (https://github.com/choderalab/pymbar).

## 3 Results and discussion

First, the root mean square deviation (RMSD) of the protein backbone on the SAM surface is analyzed (Figure S1 in the Supplementary Information). The RMSD results show that the simulations have converged. Meanwhile, the secondary structure analysis provided in Figure S2 of the Supplementary Information reveals that the primary secondary structure of protein has been retained, indicating that the native conformation has not been disrupted.

The binding affinity of protein to the SAM surface in this study is characterized by the number of hydrogen bonds formed between protein and the SAM, the difference in alchemical free energy, and the contact area between them. The average number of hydrogen bonds is listed in Table 1, which represents the number of hydrogen bonds among OH groups of SAM, between the SAM and interfacial water, and between the SAM and protein at various packing densities. The alchemical free energy differences of protein on single or double OH-terminated SAMs are shown in Figure 7 at the three packing densities. The contact area of protein on the SAM is shown in Figure 8, which is calculated using the following formula[36-38]

$$\text{contact area} = \frac{1}{2} \left( \left( \text{SASA}_{\text{protein}} + \text{SASA}_{\text{surface}} \right) - \text{SASA}_{\text{complex}} \right) \quad (1)$$

where $\text{SASA}_{\text{protein}}$ and $\text{SASA}_{\text{surface}}$ represent the solvent accessible surface area of the isolated protein and the SAM, respectively, and $\text{SASA}_{\text{complex}}$ is that of the protein and SAM complex. A probe of 1.4 Å radius is employed to calculate the individual solvent accessible surface area of the protein, the SAM, and the complex.

### 3.1 The existence of interfacial water

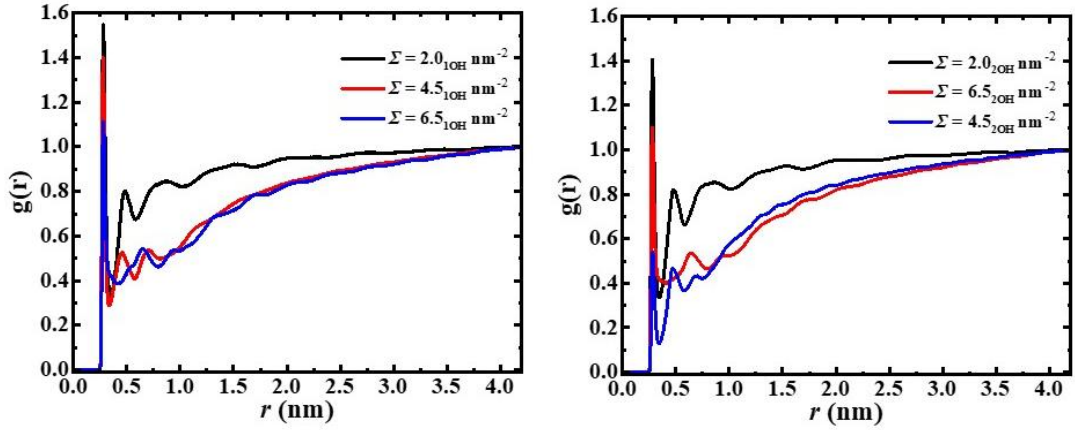

Figure 2. Radial Distribution Function (RDF) of oxygen atoms in water with respect to the topmost OH groups of the SAM at the three packing densities. The left panel corresponds to the case of SAM terminated with a single OH group (denoted by the subscript 1OH, same as below), while the right panel corresponds to the case of SAM terminated with two OH groups (denoted by the subscript 2OH, same as below).

The binding affinity of protein to the SAM surface is influenced by multiple factors, including the well-known physical barrier effect of interfacial water.[13] Due to this barrier effect, direct adsorption of protein onto the SAM surface is hindered, resulting in OH-SAM exhibiting resistance to protein adsorption. Therefore, we analyzed RDF of oxygen atoms in water relative to the topmost OH groups of the SAM to reveal the presence of interfacial water. As shown in Figure 2, the first peak of RDF reflects the presence of interfacial water near the SAM at the three packing densities. We observe that, irrespective of the single or double OH-terminated SAM systems, the peak of RDF at $\Sigma = 2.0$ nm$^{-2}$ is consistently the highest, while the RDF at $\Sigma = 4.5_{2OH}$ nm$^{-2}$ exhibits the lowest peak among all six systems. It is worth noting that the peak of the RDF at $\Sigma = 6.5_{1OH}$ nm$^{-2}$ is lower than that at $\Sigma = 4.5_{1OH}$ nm$^{-2}$, contrary to the order of peak heights for water density (Figure 4) at the two packing densities. This indicates that interfacial water at $\Sigma = 6.5_{1OH}$ nm$^{-2}$ is denser, analogous to the situation observed at $\Sigma = 2.0$ nm$^{-2}$. The RDF analysis confirms the presence of interfacial water on these SAM surfaces. The barrier effect of interfacial water may have varying degrees of impact on the resistance toward protein. Here, we define the water within 6.5 Å above the topmost OH groups of the SAM as interfacial water.

**3.2 The effect of compactness and order of interfacial water**

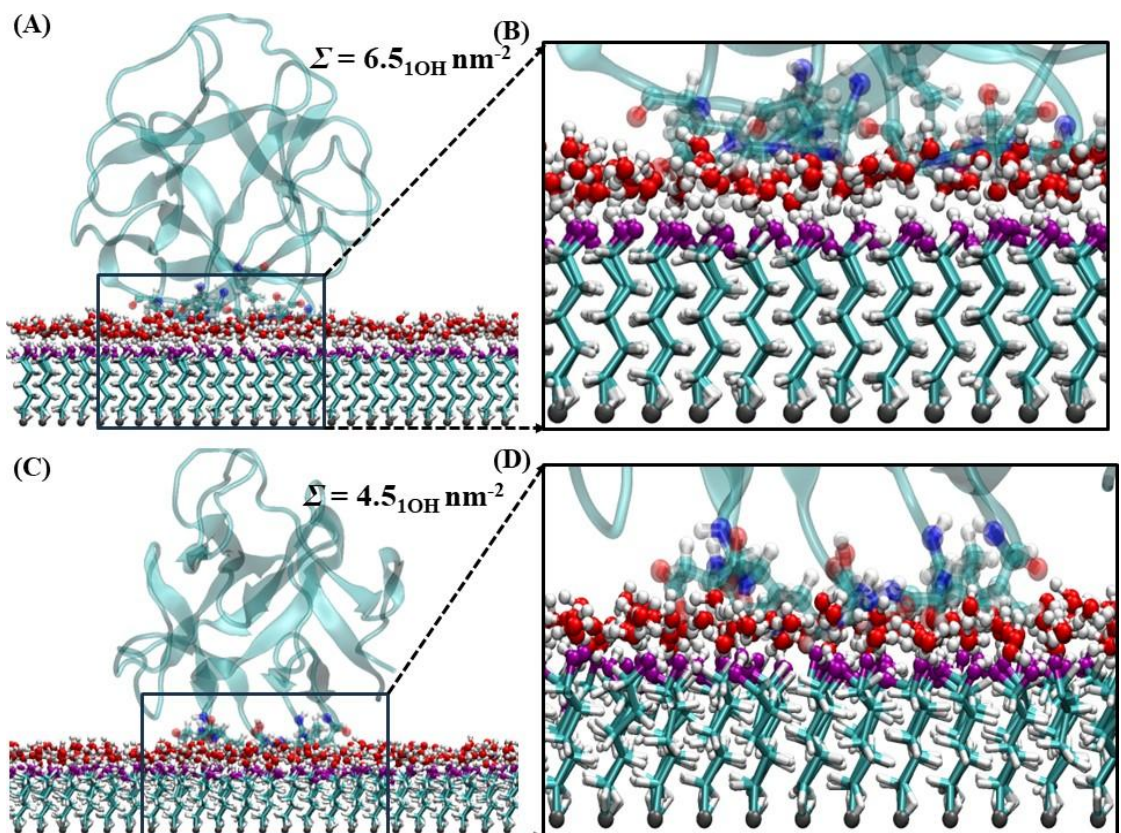

Figure 3. Snapshots of protein adsorption on the SAM at $\Sigma = 6.5_{1OH}$ nm$^{-2}$ (A, B) and $4.5_{1OH}$ nm$^{-2}$ (C, D), with interfacial water highlighted separately. The rest of the system is omitted for clarity. The subfigure represents a partial enlargement of interfacial water on the SAM surface. The proteins are represented using a NewCartoon method with a transparent mode. The protein residues near the SAM surface are displayed in CPK mode. Water molecules are displayed in CPK mode with red and white spheres. The color codes and atomic representations for the rest are the same as those in Figure 1.

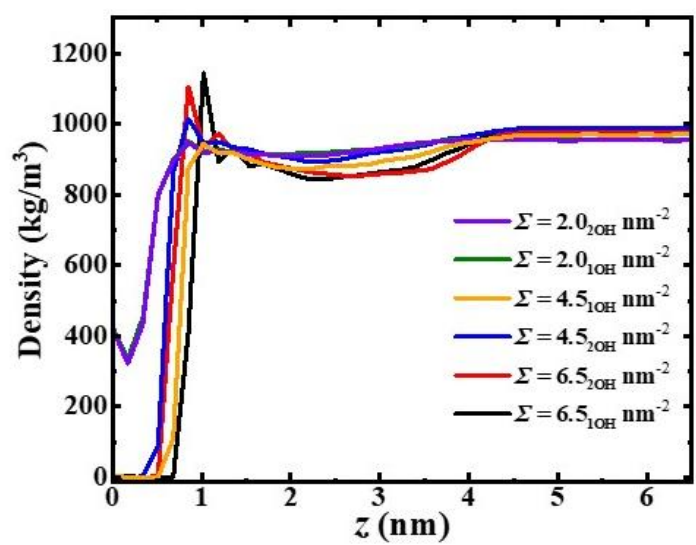

Figure 4. Density distribution of water molecules along the *z*-axis (perpendicular to the SAM surface) at various packing densities. The two density curves at $\Sigma = 2.0$ nm$^{-2}$ exhibit significant overlap.

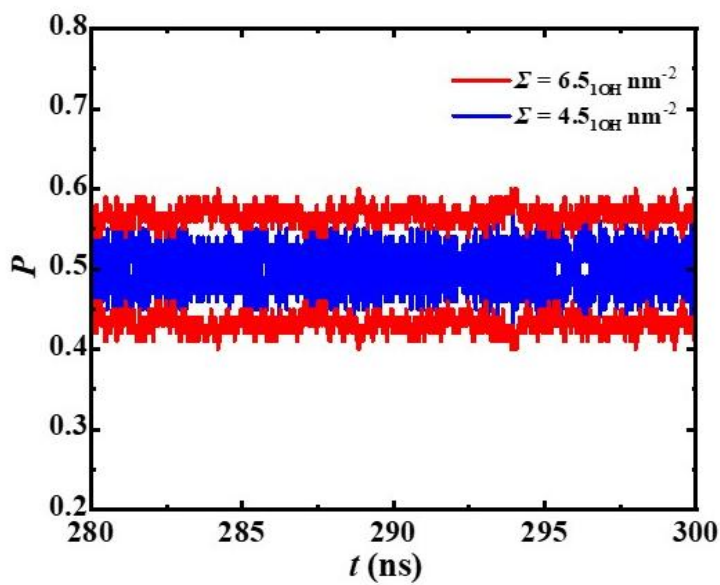

Figure 5. Probability distribution of dipole orientation of interfacial water as a function of time. The red color corresponds to the case of $\Sigma = 6.5_{1OH}$ nm$^{-2}$, while the blue color represents that at $\Sigma = 4.5_{1OH}$ nm$^{-2}$. There are two curves of the same color for each packing density, where the upper curve represents the upward dipole orientation of interfacial water (i.e., the oxygen atom of water facing downward), and the lower curve represents the downward dipole orientation of interfacial water (i.e., the oxygen atom of water facing upward). Here, we define the dipole orientation of interfacial water as an angle less than 90° with the *z*-axis, indicating an upward dipole orientation. Otherwise, downward.

Table 1. Hydrogen bond number analysis

| $N_{OH}$[a] | $\Sigma$ (nm$^{-2}$) | SAM-SAM | SAM-water | SAM-protein |
|---|---|---|---|---|
| 1OH | 2.0 | 0.14 | 0.92 | 4.39 |
|  | 4.5 | 0.85 | 6.19 | 3.92 |
|  | 6.5 | 1.32 | 7.27 | 3.35 |
| 2OH | 2.0 | 0.30 | 1.71 | 5.82 |
|  | 4.5 | 8.02 | 2.85 | 0.64 |
|  | 6.5 | 2.08 | 6.98 | 5.56 |

[a] The number of OH groups in the alkyl chain. 1OH represents the single OH-terminated SAM system, while 2OH represents the double OH-terminated SAM system.

As mentioned earlier, the barrier effect of interfacial water reduces protein adsorption on the SAM, and we further found that the resistance of SAM towards protein exhibits variations under the influence of interfacial water at $\Sigma = 4.5_{1OH}$ nm$^{-2}$ and $6.5_{1OH}$ nm$^{-2}$, as shown in Figure 3. It was found that a lower number of hydrogen bonds between protein and the SAM at $\Sigma = 6.5_{1OH}$ nm$^{-2}$ (Table 1), indicating a weaker binding affinity of protein to the SAM compared to that at $\Sigma = 4.5_{1OH}$ nm$^{-2}$. Furthermore, the difference in alchemical free energy at $\Sigma = 6.5_{1OH}$ nm$^{-2}$ is greater compared to that at $\Sigma = 4.5_{1OH}$ nm$^{-2}$ (Figure 7), indicating a less stable binding of protein to the SAM at $\Sigma = 6.5_{1OH}$ nm$^{-2}$. This further reflects the weaker binding affinity of protein to the SAM at $\Sigma = 6.5_{1OH}$ nm$^{-2}$. In Figure 3, we notice that the arrangement of OH groups of SAM at $\Sigma = 6.5_{1OH}$ nm$^{-2}$ is more regular and ordered, consequently leading to a more compact and ordered distribution of interfacial water on the SAM surface. We speculate that this is responsible for the weaker binding affinity of protein to the SAM at $\Sigma = 6.5_{1OH}$ nm$^{-2}$. To demonstrate this, we first analyze the density of water molecules perpendicular to the SAM direction. As shown in Figure 4, the density of water molecules at $\Sigma = 6.5_{1OH}$ nm$^{-2}$ is significantly higher than that at $\Sigma = 4.5_{1OH}$ nm$^{-2}$, demonstrating the compactness of interfacial water. Further, we analyze the probability distribution of the dipole orientation[39-42] of interfacial water to demonstrate the order of interfacial water at $\Sigma = 6.5_{1OH}$ nm$^{-2}$, as shown in Figure 5. The two blue curves at $\Sigma = 4.5_{1OH}$ nm$^{-2}$ are located between the two red curves at $\Sigma = 6.5_{1OH}$ nm$^{-2}$, indicating that a more disordered dipole orientation of interfacial water caused by the disorder in the distribution of OH groups of SAM at $\Sigma = 4.5_{1OH}$ nm$^{-2}$, i.e., the oxygen atoms of interfacial water exhibit a random distribution, with a mixture of upward and downward orientations. In contrast, the more regular and ordered distribution of OH groups of SAM at $\Sigma = 6.5_{1OH}$ nm$^{-2}$ results in a more ordered dipole orientation of interfacial water compared to that at $\Sigma = 4.5_{1OH}$ nm$^{-2}$. The upper red curve at $\Sigma = 6.5_{1OH}$ nm$^{-2}$ indicates that more oxygen atoms of interfacial water are facing downward to form hydrogen bonds with the upward hydrogen atoms of OH groups of SAM. In Table 1, the higher number of hydrogen bonds (7.27 nm$^{-2}$) between the SAM and interfacial water at $\Sigma = 6.5_{1OH}$ nm$^{-2}$ further confirms a more favorable structure arrangement for hydrogen bond formation between the SAM and interfacial water compared to that (6.19 nm$^{-2}$) at $\Sigma = 4.5_{1OH}$ nm$^{-2}$. Therefore, the compactness and order of interfacial water enhance its barrier effect, thereby enhancing resistance to protein adsorption on the SAM.

**3.3 The spatial hindrance effect of SAM**

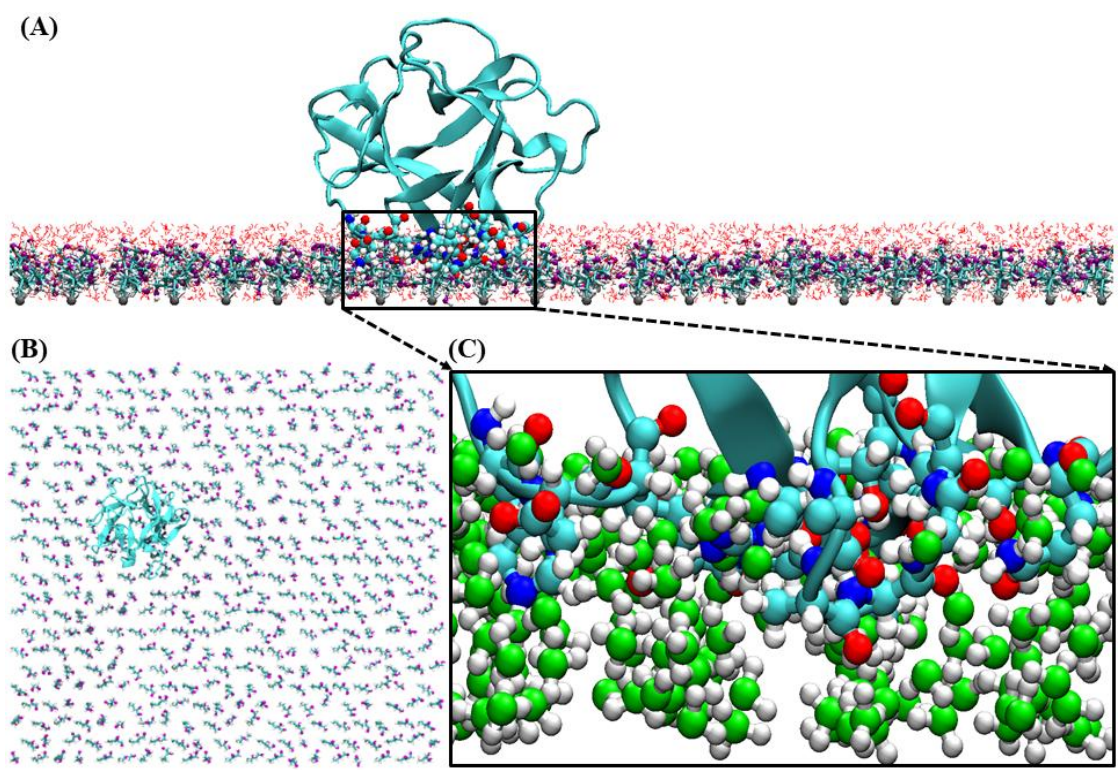

Figure 6. (A) A side view of protein embedded into SAM at $\Sigma = 2.0_{2OH}$ nm$^{-2}$. The interfacial water and the water penetrating into SAM are highlighted separately in a linear mode. The protein is represented in a NewCartoon mode, while the protein residues near the SAM surface are displayed using the CPK representation. The rest of the system is omitted for clarity. The color codes and atomic representations for the rest are the same as those in Figure 1. (B) A top view of protein embedded into SAM. Water molecules are also omitted for clarity. (C) A magnified view of protein embedded into SAM. Water molecules are represented by green and white spheres.

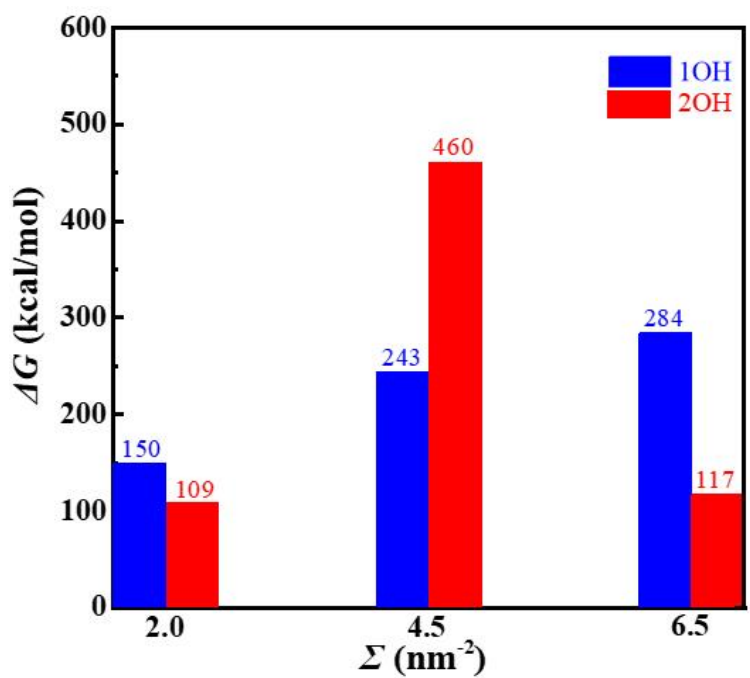

Figure 7. Free energy differences of proteins on single or double OH-terminated SAMs at various packing densities. The blue color represents the free energy difference for the single OH-terminated SAM system, while the red color represents the free energy difference for the double OH-terminated SAM system. The data is divided into three groups, i.e., the first two are grouped together, representing the free energy difference at $\Sigma = 2.0$ nm$^{-2}$. The middle two are grouped together, representing the free energy difference at $\Sigma = 4.5$ nm$^{-2}$. The last two are grouped together, representing the free energy difference at $\Sigma = 6.5$ nm$^{-2}$.

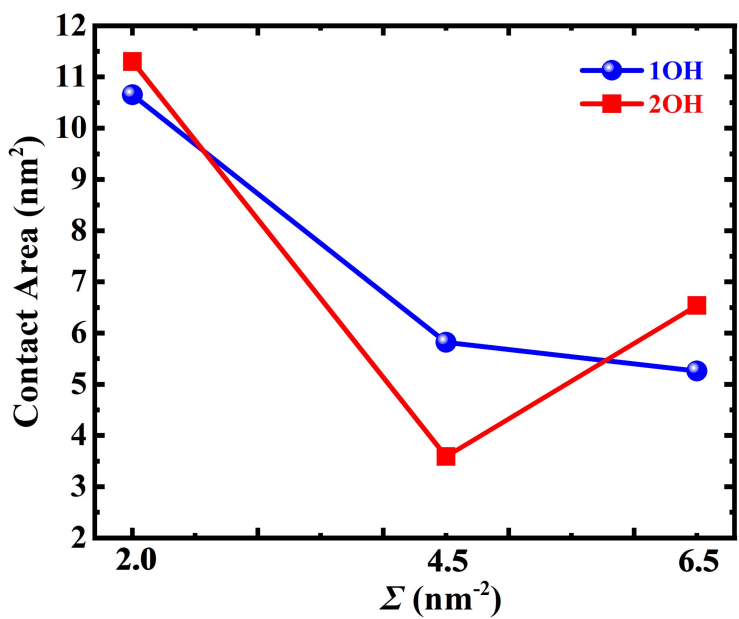

Figure 8. The average contact area of proteins on single or double OH-terminated SAM surfaces at the three packing densities. The blue color represents the contact area of protein on the single OH-terminated SAM, while the red color represents the contact area of protein on the double OH-terminated SAM.

The binding affinity of protein to the SAM surface is also influenced by the spatial hindrance effect of SAM. From Figure 6, it can be observed that the protein is embedded into SAM, which can be primarily ascribed to the weak spatial hindrance effect of SAM at $\Sigma = 2.0_{2OH}$ nm$^{-2}$. Meanwhile, the spatial hindrance effect leads to the lowest number of hydrogen bonds among OH groups of the SAM (Table 1, 0.14 for $\Sigma = 2.0_{1OH}$ nm$^{-2}$ and 0.30 for $\Sigma = 2.0_{2OH}$ nm$^{-2}$) within the three packing densities, indicating the minimal impact of the structure of SAM itself (detailed analysis in Section 3.5) on resistance towards protein. Furthermore, the weak spatial hindrance effect also results in the lowest number of hydrogen bonds between SAM and interfacial water at $\Sigma = 2.0$ nm$^{-2}$ (Table 1, 0.92 for $\Sigma = 2.0_{1OH}$ nm$^{-2}$ and 1.71 for $\Sigma = 2.0_{2OH}$ nm$^{-2}$), indicating the weakest barrier effect of interfacial water on resistance towards protein. As shown in Figure 6, we notice that the weak spatial hindrance effect allows water penetration into SAM, which can be reflected by the non-zero minimum density at $\Sigma = 2.0$ nm$^{-2}$ in Figure 4. Furthermore, the RDF at $\Sigma = 2.0$ nm$^{-2}$ exhibits the highest peak, indicating a substantial accumulation of water molecules near the SAM. Also, the relatively low density of water in Figure 4 further confirms the weak barrier effect of interfacial water, which is consistent with the lowest number of hydrogen bonds between the SAM and interfacial water at $\varSigma = 2.0$ nm$^{-2}$. Therefore, protein can easily displace water molecules and subsequently embed itself into SAM. At this time, the number of hydrogen bonds (Table 1, 4.39 for $\varSigma = 2.0_{1OH}$ nm$^{-2}$ and 5.82 for $\varSigma = 2.0_{2OH}$ nm$^{-2}$) between the protein and SAM is the largest, indicating the strongest binding affinity of protein to the SAM. We observe that the difference in alchemical free energy of protein on the SAM surface was minimized (Figure 7) at $\varSigma = 2.0$ nm$^{-2}$, corresponding to the largest contact area between protein and the SAM (Figure 8). These findings further confirm the strongest binding affinity of protein to the SAM. Hence, the protein is embedded into SAM under the influence of the weak spatial hindrance effect, and the SAM exhibits a lack of resistance towards the protein. The simulation snapshots at $\varSigma = 2.0_{1OH}$ nm$^{-2}$ are provided in Figure S4 of Supplementary Information.

**3.4 The effect of the number of hydroxyl groups per unit area of SAM**

We observe that the resistance of SAM towards protein is also modulated by the number of hydroxyl groups per unit area of SAM. As shown in Table 1, the number of hydrogen bonds between protein and the SAM at $\varSigma = 6.5_{2OH}$ nm$^{-2}$ (5.56) is greater than that at $\varSigma = 6.5_{1OH}$ nm$^{-2}$ (3.35) and $\varSigma = 4.5_{1OH}$ nm$^{-2}$ (3.92), indicating a weaker resistance of SAM towards protein at $\varSigma = 6.5_{2OH}$ nm$^{-2}$. It can be seen that the number of hydrogen bonds (0.85 at $\varSigma = 4.5_{1OH}$ nm$^{-2}$, 1.32 at $\varSigma = 6.5_{1OH}$ nm$^{-2}$, and 2.08 at $\varSigma = 6.5_{2OH}$ nm$^{-2}$) formed among OH groups of SAM in the three cases is significantly lower than the number of hydrogen bonds (6.19 at $\varSigma = 4.5_{1OH}$ nm$^{-2}$, 7.27 at $\varSigma = 6.5_{1OH}$ nm$^{-2}$, and 6.98 at $\varSigma = 6.5_{2OH}$ nm$^{-2}$) formed between SAM and interfacial water, indicating that the barrier effect of interfacial water plays a dominant role in the resistance of SAM towards protein. Furthermore, we notice that the number of OH groups per unit area of double OH-terminated SAM differs significantly from that of single OH-terminated SAM, i.e., the number of OH groups (12.95) per unit area of SAM at $\varSigma = 6.5_{2OH}$ nm$^{-2}$ is approximately 2 to 3 times that at $\varSigma = 6.5_{1OH}$ nm$^{-2}$ (6.56) and $\varSigma = 4.5_{1OH}$ nm$^{-2}$ (4.59). The SAM at $\varSigma = 6.5_{2OH}$ nm$^{-2}$ can provide more hydrogen bonding sites to form hydrogen bonds with protein, and thus result in the highest number of hydrogen bonds formed between the SAM and protein in the three cases, indicating that the SAM at $\varSigma = 6.5_{2OH}$ nm$^{-2}$ exhibits the weakest resistance to protein adsorption. Furthermore, it can be seen in Figure 7 that the alchemical free energy difference of protein adsorption on the SAM at $\varSigma = 6.5_{2OH}$ nm$^{-2}$ is lower than that at the other two packing densities, further confirming the weakest resistance of SAM at $\varSigma = 6.5_{2OH}$ nm$^{-2}$ to protein adsorption in the three cases. Therefore, the number of hydroxyl groups per unit area of SAM can significantly impact the resistance of SAM to protein adsorption.

**3.5 The impact of the structure of SAM itself**

We further observe that the resistance of SAM towards protein is also affected by the structure of SAM itself. As shown in Table 1, the number of hydrogen bonds (0.64) between protein and the SAM at $\Sigma = 4.5_{2OH}$ nm$^{-2}$ (the simulation snapshot is provided in Figure S5 of Supplementary Information) is the lowest among all systems, indicating the strongest resistance of SAM to protein adsorption. We notice that the number of hydrogen bonds (2.85) between the SAM and interfacial water at $\Sigma = 4.5_{2OH}$ nm$^{-2}$ is the lowest at all medium and high packing densities, while the number of hydrogen bonds (8.02) among OH groups of SAM is the highest. This trend at $\Sigma = 4.5_{2OH}$ nm$^{-2}$ is opposite to the corresponding number of hydrogen bonds (i.e., SAM-water, SAM-SAM) at other medium and high packing densities, indicating that the effect of the structure of SAM itself on resistance to protein adsorption is significantly greater than that of interfacial water. Upon careful examination of the SAM structure, it is found that a nearly perfect ice-like hydrogen bonding structure (the simulation snapshot is provided in Figure S6 of Supplementary Information) is formed among OH groups of SAM at $\Sigma = 4.5_{2OH}$ nm$^{-2}$. Further, the hydrogen bonding sites among OH groups of SAM are close to saturation, leaving only a small number of hydrogen bond sites available for binding with protein. This results in the lowest number of hydrogen bonds formed between the SAM and protein, corresponding to the largest difference in the alchemical free energy of protein on the SAM (Figure 7) and the smallest contact area between them (Figure 8). Therefore, the SAM exhibits the strongest resistance to protein adsorption due to the structure of SAM itself.

## 4 Conclusions

In this study, we have utilized MD simulations and alchemical free energy calculations to systematically investigate the factors influencing the resistance to protein adsorption on single or double OH-terminated SAMs at various packing densities. The results show that four factors play a significant role in the resistance of SAM to protein adsorption, namely, the compactness and order of interfacial water, the spatial hindrance effect of SAM, the number of hydroxyl groups per unit area of SAM, and the structure of SAM itself.

For the first time, we found that interfacial water at $\Sigma = 6.5_{1OH}$ nm$^{-2}$ is denser and more ordered compared to that at $\Sigma = 4.5_{1OH}$ nm$^{-2}$, resulting in a lower number of hydrogen bonds between the SAM and protein at $\Sigma = 6.5_{1OH}$ nm$^{-2}$. Therefore, the compactness and order of interfacial water enhance its barrier effect, thereby enhancing the resistance of SAM towards protein. We further found that the protein is embedded into SAM due to the weak spatial hindrance effect of SAM at $\Sigma = 2.0$ nm$^{-2}$, corresponding to the maximum number of hydrogen bonds between protein and the SAM, and the SAM exhibits a lack of resistance towards protein. We also found that the number of hydroxyl groups per unit area of SAM at $\Sigma = 6.5_{2OH}$ nm$^{-2}$ is approximately 2 to 3 times that at $\Sigma = 6.5_{1OH}$ nm$^{-2}$ and $\Sigma = 4.5_{1OH}$ nm$^{-2}$. The SAM at $\Sigma = 6.5_{2OH}$ nm$^{-2}$ can provide more hydrogen bonding sites to form hydrogen bonds with protein, resulting in a weaker resistance of SAM to protein adsorption. Therefore, the number of OH groups per unit area of SAM can significantly affect the resistance of SAM towards protein. Meanwhile, we also found that the structure of SAM itself, i.e., the formation of a nearly perfect ice-like hydrogen bond structure results in the least number of hydrogen bonds formed between the SAM and protein, and the SAM exhibits the strongest resistance towards protein.

Our findings enrich and improve the understanding of the mechanism underlying the resistance of OH-SAM to protein adsorption, particularly the traditional barrier effect of interfacial water. These findings are of great significance for the rational design and precise customization of OH-terminated anti-fouling surfaces, with potential applications in ship navigation and medical implants.

**Conflicts of interest**

There are no conflicts to declare.

**Acknowledgements**

This work was supported by the National Natural Science Foundation of China (Grants No. 12075201), the Science and Technology Planning Project of Jiangsu Province (BK20201428), the Postgraduate Research & Practice Innovation Program of Jiangsu Province (KYCX21_3193), and the Special Program for Applied Research on Supercomputation of the NSFC-Guangdong Joint Fund (the second phase).

**References**

1. A. Jain and N. B. Bhosle, *Biofouling*, 2009, **25**, 13-19.
2. C. Ma, H. Yang, X. Zhou, B. Wu and G. Zhang, *Colloid Surface B*, 2012, **100**, 31-35.
3. X. Li, S. Li, X. Huang, Y. Chen, J. Cheng and A. Zhan, *Mar Environ Res*, 2021, **170**, 105409.
4. C. C. C. R. de Carvalho, *Front. Mar. Sci.*, 2018, **5**, 1-11.
5. M. C. L. Martins, C. Fonseca, M. A. Barbosa and B. D. Ratner, *Biomaterials*, 2003, **24**, 3697-3706.
6. S. Choi, Y. Yang and J. Chae, *Biosens. Bioelectron.*, 2008, **24**, 893-899.
7. J. Guo, P. Zhang, Y. Chen, Y. Shen, X. Hu, P. Yan, J. Yang, F. Fang, C. Li, X. Gao and G. Wang, *Chem. Eng. J.*, 2015, **279**, 516-521.
8. T. Miyake, T. Tanii, K. Kato, T. Zako, T. Funatsu and I. Ohdomari, *Nanotechnology*, 2007, **18**, 305304.
9. S. Huang, Q. Hou, D. Guo, H. Yang, T. Chen, F. Liu, G. Hu, M. Zhang, J. Zhang and J. Wang, *RSC. Adv.*, 2017, **7**, 39530-39538.
10. J. Zhao, Q. Wang, G. Liang and J. Zheng, *Langmuir*, 2011, **27**, 14876-14887.

11. Q. Wang, C. Zhao, J. Zhao, J. Wang, J. C. Yang, X. Yu and J. Zheng, *Langmuir*, 2010, **26**, 3308-3316.
12. C. Peng, J. Liu, D. Zhao and J. Zhou, *Langmuir*, 2014, **30**, 11401-11411.
13. Q. Wang, J. Zhao, X. Yu, C. Zhao, L. Li and J. Zheng, *Langmuir*, 2010, **26**, 12722-12732.
14. D. Mao, X. Wang, Y. Wu, Z. Gu, C. Wang and Y. Tu, *Nanoscale*, 2021, **13**, 19604-19609.
15. D. Mao, Y. Y. Wu and Y. Tu, *Phys. Chem. Chem. Phys.*, 2023, **25**, 21376-21382.
16. P. Guo, Y. Tu, J. Yang, C. Wang, N. Sheng and H. Fang, *Phys. Rev. Lett.*, 2015, **115**, 186101.
17. Z. Chen, C. Qi, X. Teng, B. Zhou and C. Wang, *Commun. Theor. Phys.*, 2021, **73**, 115501.
18. W. Miao, Y. Tian and L. Jiang, *Acc. Chem. Res.*, 2022, **55**, 1467-1479.
19. Y. Tian and L. Jiang, *Nat. Mater.*, 2013, **12**, 291-292.
20. M. Jakob, J. Lubkowski, B. R. O'Keefe and A. Wlodawer, *Acta Crystallogr F*, 2015, **71**, 1429-1436.
21. C. He, H. Zhang, C. Lin, L. Wang and S. Yuan, *Chem. Phys. Lett.*, 2017, **676**, 144-149.
22. X. Fan, Y. Fang, W. Zhou, L. Yan, Y. Xu, H. Zhu and H. Liu, *Mater. Horiz.*, 2021, **8**, 997-1007.
23. X. Ou, B. Xue, Y. Lao, Y. Wutthinitikornkit, R. Tian, A. Zou, L. Yang, W. Wang, Y. Cao and J. Li, *Sci. Adv.*, 2020, **6**, eabb7620.
24. H. J. C. Berendsen, J. R. Grigera and T. P. Straatsma, *J. Phys. Chem.*, 1987, **91**, 6269-6271.
25. D. Van Der Spoel, E. Lindahl, B. Hess, G. Groenhof, A. E. Mark and H. J. Berendsen, *J. Comput. Chem.*, 2005, **26**, 1701-1718.
26. M. J. Abraham, T. Murtola, R. Schulz, S. Páll, J. C. Smith, B. Hess and E. Lindahl, *SoftwareX*, 2015, **1-2**, 19-25.
27. W. L. Jorgensen, D. S. Maxwell and J. Tirado-Rives, *J. Am. Chem. Soc.*, 1996, **118**, 11225-11236.
28. T. Darden, D. York and L. Pedersen, *J. Chem. Phys.*, 1993, **98**, 10089-10092.
29. U. Essmann, L. Perera, M. L. Berkowitz, T. Darden, H. Lee and L. G. Pedersen, *J. Chem. Phys.*, 1995, **103**, 8577-8593.
30. B. Hess, H. Bekker, H. J. C. Berendsen and J. G. E. M. Fraaije, *J. Comput. Chem.*, 1997, **18**, 1463-1472.
31. M. Aldeghi, A. Heifetz, M. J. Bodkin, S. Knapp and P. C. Biggin, *Chem. Sci.*, 2016, **7**, 207-218.
32. D. L. Mobley, J. D. Chodera and K. A. Dill, *J. Chem. Phys.*, 2006, **125**, 084902.
33. T. C. Beutler, A. E. Mark, R. C. van Schaik, P. R. Gerber and W. F. van Gunsteren, *Chem. Phys. Lett.*, 1994, **222**, 529-539.
34. R. W. Pastor, B. R. Brooks and A. Szabo, *Mol. Phys.*, 2006, **65**, 1409-1419.
35. M. R. Shirts and J. D. Chodera, *J. Chem. Phys.*, 2008, **129**, 124105.
36. B. Luan, T. Huynh, L. Zhao and R. Zhou, *ACS nano*, 2015, **9**, 663-669.
37. R. Ye, W. Song, M. Feng and R. Zhou, *Nanoscale*, 2021, **13**, 19255-19263.

38. M. Feng, H. Kang, Z. Yang, B. Luan and R. Zhou, *J. Chem. Phys.*, 2016, **144**, 225102.
39. D. Mao, X. Wang, G. Zhou, S. Zeng, L. Chen, J. Chen and C. Dai, *Chin. Phys. Lett.*, 2019, **36**.
40. Y. Tu, H. Lu, Y. Zhang, T. Huynh and R. Zhou, *J. Chem. Phys.*, 2013, **138**, 015104.
41. Y. Tu, R. Zhou and H. Fang, *Nanoscale*, 2010, **2**, 1976-1983.
42. Y. Tu, P. Xiu, R. Wan, J. Hu, R. Zhou and H. Fang, *Proc. Natl. Acad. Sci. U. S. A.*, 2009, **106**, 18120-18124.